\documentclass[letterpaper,conference]{IEEEtran}
\usepackage{graphicx}
\usepackage{url}

\usepackage{atbegshi,picture}
\usepackage{lipsum}

\AtBeginShipout{\AtBeginShipoutUpperLeft{%
  \put(\dimexpr\paperwidth-1cm\relax,-1.5cm){\makebox[0pt][r]{\framebox{IEEE 13th International Conference on Internet of Things (iThings 2020), November, 2020}}}%
}}

\begin{document}
\title{Security, Privacy and Ethical Concerns of IoT Implementations in Hospitality Domain\thanks{This work is supported by a grant from  Cisco/Silicon Valley Foundation.}}
%
%

\author{Suat Mercan\inst{1}\and
Kemal Akkaya\inst{1}\and
Lisa Cain\inst{2}\and
John Thomas\inst{2}}
%
%

\author{\IEEEauthorblockN{Suat Mercan and Kemal Akkaya} 
\IEEEauthorblockA{Dept. of Electrical and Computer Engineering\\
Florida International University\\
Miami, FL 33174\\ 
Email: \{smercan, kakkaya\}@fiu.edu}
}

\author{\IEEEauthorblockN{Suat Mercan\IEEEauthorrefmark{1},
Kemal Akkaya\IEEEauthorrefmark{1}, Lisa Cain\IEEEauthorrefmark{2}, and John Thomas\IEEEauthorrefmark{2}} 
\IEEEauthorblockA{\IEEEauthorrefmark{1}Dept. of Elec. and Comp. Engineering, Florida International University, Miami, FL 33174\\ Email: \{smercan,kakkaya\}@fiu.edu}
\IEEEauthorblockA{\IEEEauthorrefmark{2}Chaplin School of Hospitality \& Tourism Management, Florida International University, Miami, FL 33181, USA\\Email: \{lcain,thomasj\}@fiu.edu}
}



%
\maketitle              
\begin{abstract}

The Internet of Things (IoT) has been on the rise in the last decade as it finds applications in various domains. 
Hospitality is one of the pioneer sectors that has adopted this technology to create novel services such as smart hotel rooms, personalized services etc. Hotels, restaurants, theme parks, and cruise ships are some specific application areas to improve customer satisfaction by creating an intense interactive environment and data collection with the use of appropriate sensors and actuators. However, applying IoT solutions in the hospitality environment has some unique challenges such as easy physical access to devices. 
In addition, due to the very nature of these domains, the customers are at the epicenter of these IoT technologies that result in a massive amount of data collection from them. Such data and its management along with business purposes also raises new concerns regarding privacy and ethical considerations. Therefore, this paper surveys and analyzes security, privacy and ethical issues regarding the utilization of IoT devices by focusing on the hospitality industry specifically. We explore some exemplary uses, cases, potential problems and solutions in order to contribute to better understanding and guiding the business operators in this sector.

\end{abstract}

\begin{IEEEkeywords}
Internet of Things, Hospitality, Security, Privacy, Ethics
\end{IEEEkeywords}

\section{Introduction}

The Internet of Things (IoT) provides great opportunities to improve hospitality businesses in numerous ways, mainly by increasing operational efficiency and customer satisfaction with a vast amount of data collection\cite{car2019internet,pelet2019internet}. It brings new revenue streams by adding value to services via IoT, which facilitates integration of the physical and virtual world. Smart room is a well known example implemented by hoteliers that provides great comfort to guest in controlling items in the room remotely. Smart-lock enables a guest to check-in without a receptionist.
Context-aware and personalized services are main characteristics of IoT-based service, which necessitates the collection of personal data \cite{xiang2017big}. In a typical IoT system, data is collected from the user and transmitted using various protocols to a cloud server, which processes the data to generate an action \cite{ray2016survey}. Moreover, the devices could be integrated with the rest of the system to provide real-time information to automate and monitor the operations. A guest within the perimeters of a hotel, theme park, cruiseship is tracked using a wearable device, and he is offered promotional activities, meals based on his preferences while the crowd is managed smoothly.

In this aspect, there are many vulnerabilities and risks in the overall system that could be exploited by malicious actors against system security and customer privacy \cite{sicari2015security}. While security is the primary concern for business operators (e.g., hotel lockdown), privacy violation by business operators or an attacker is of pivotal concern for end-users. These issues remain as the primary impediment for societal acceptance of IoT.  
A hospitality environment implementing IoT solutions is exposed to various attacks. The perpetrators may target IoT devices or utilize these devices to disrupt the operation of the hotel such as the case of Romantik Seehotel in Austria whose locking system was frozen by ransomware \cite{romantikhotel}. The hotel had to pay \$1800 (two bitcoins) to end this Distributed Denial of Service (DDoS) attack. If the IoT devices are integrated with other information systems, perpetrators may access sensitive information such as credit card numbers or financial records \cite{fishtank}. Thus, it is important to develop and employ security measures to prevent such incidents that may cause thousands to millions of dollars to businesses. However, the nature of the IoT devices (e.g., resource scarcity) makes it harder to apply traditional security techniques. Data encryption using digital keys and certificates require intensive computations and key management schemes, which are not feasible in an IoT environment. Moreover, physical attacks could be easier to achieve in a hotel environment where access to these devices are not restricted. Any guest can infiltrate the system by exploiting physical proximity as most of the devices are not protected electronically. Furthermore, the sheer increase in the number and variety of devices will bring additional targets vulnerable to attacks. More items and applications mean more attack surfaces \cite{hossain2015towards}. 

IoT enables computers to observe, identify and understand the world without relying on human-entered data \cite{ashton2009internet}. Data collection, aggregation and analysis are key to success for improved customer satisfaction. Hospitality industry, in particular, needs personal information to better serve the customers, thus privacy is a major concern in this domain. Many examples exist from hotels to theme parks that adopt this technology which constantly observes the physical environment, interacts with the customers, and gathers data.  A person within the facility is tracked throughout the day and all his activities are logged to provide personalized offerings. The collected data can reveal behavioural patterns such as eating habits and fitness activities \cite{kroger2018unexpected}. This type of personal information can be used by the business owner or third parties if accessed for price discrimination, denial of insurance, and marketing purposes against the uses' will to name a few. Non-identifying data could be used to profile a customer, and even be merged with other data set to identify a person \cite{ziegeldorf2014privacy}. Data minimization and anonymization are some methods that can be adopted to reduce the risk of privacy violation.  

Pervasive data collection about customers through such devices shifts the control of information from user to business owner as some ethical concerns arise \cite{allhoff2018internet}. For instance, even though the customers are required to give consent before being exposed to this environment, they may not understand and predict the full potential consequences of information sharing. It is even challenging to make clear the type of information that will be collected from customers because of the variety of devices.  Moreover, the privacy expectation of people might differ. Thus, alternative models to 'click-and-agree' consent should be developed to allow people to choose from different level of services. A person may want to change the privacy setting from time to time. However, each device interacts with customers differently and mostly lacks the interface to ask user consent or change privacy settings \cite{allhoff2018internet}. While the data transfer is seamless, the operator must behave responsibly and not use the data beyond the consumer's intent. They also must be transparent about how the accumulated data will be used and how long it will be stored in the database. 

Despite the aforementioned concerns, business owners do not derail from the adoption of this technology because of the potential benefits. In this paper, we introduce various sectors in hospitality that adopted IoT with the implementation of specific applications. We, then, investigate security, privacy and ethical concerns of IoT utilization by focusing on hospitality domain. Hospitality environment poses some specific challenges in terms of secure and privacy preserving implementation. We aim to increase awareness of potential risks as IoT is being adopted at a great pace. Finally, we give some general guidelines to be adapted in the implementation and execution phases. 

This paper is organized as follows: Section II introduces IoT and its applications in the hospitality domain while Section III explores security problems. Section IV presents related privacy concerns and Section V covers the ethical aspects of the topic. In Section VI, we summarize general guidelines. Finally, Section VII concludes the paper.

\section{Background}

\subsection{Internet of Things}

\begin{figure}[ht]
    \centering
    \includegraphics[scale=0.55]{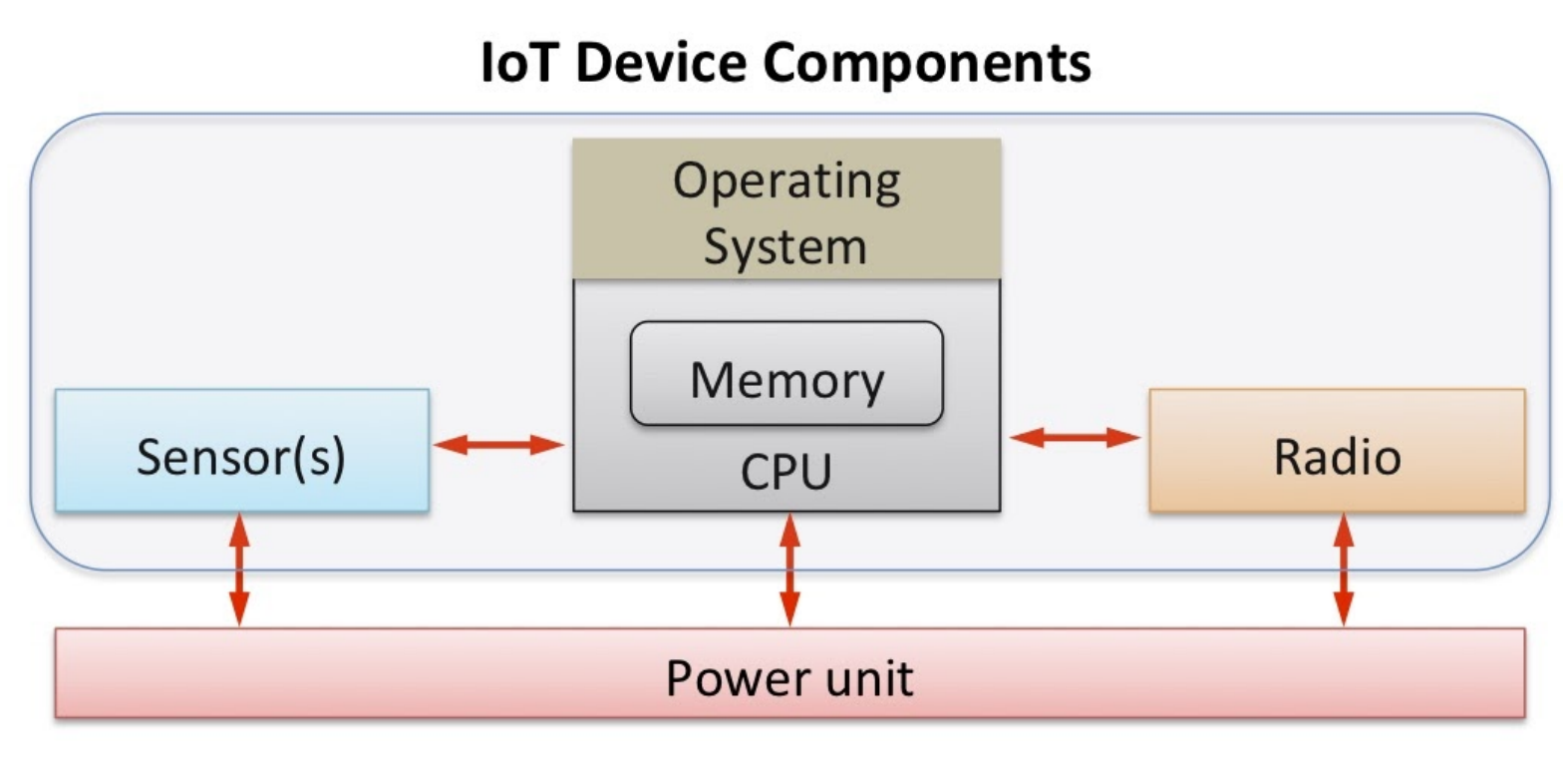}
    \caption{Components of an IoT Device.}
    \label{fig:iotcomp}
\end{figure}

IoT is perceived as an architecture that provides connectivity among people, systems and products \cite{porter2014smart}. It distinguishes from the Internet with "anytime" and "anywhere" connection capability. It consists of end-devices, a myriad of sensors, various communication links, and an intelligence center running on the cloud \cite{butun2019security,patel2016internet}. Pervasive presence of sensors enables continuous data collection with time and context information. Collected data is aggregated and analysed to generate appropriate response and action.

An IoT device as shown in Fig. \ref{fig:iotcomp} has 3 main components: 1) a primitive CPU and memory for processing and storage; 2) sensors for sensing the environment; and 3) a radio interface for communicating its data \cite{patel2016internet}. In addition to these components, the majority of the IoT devices are said to be battery-operated and small in size. 

\begin{figure}[ht]
    \centering
    \includegraphics[scale=0.65]{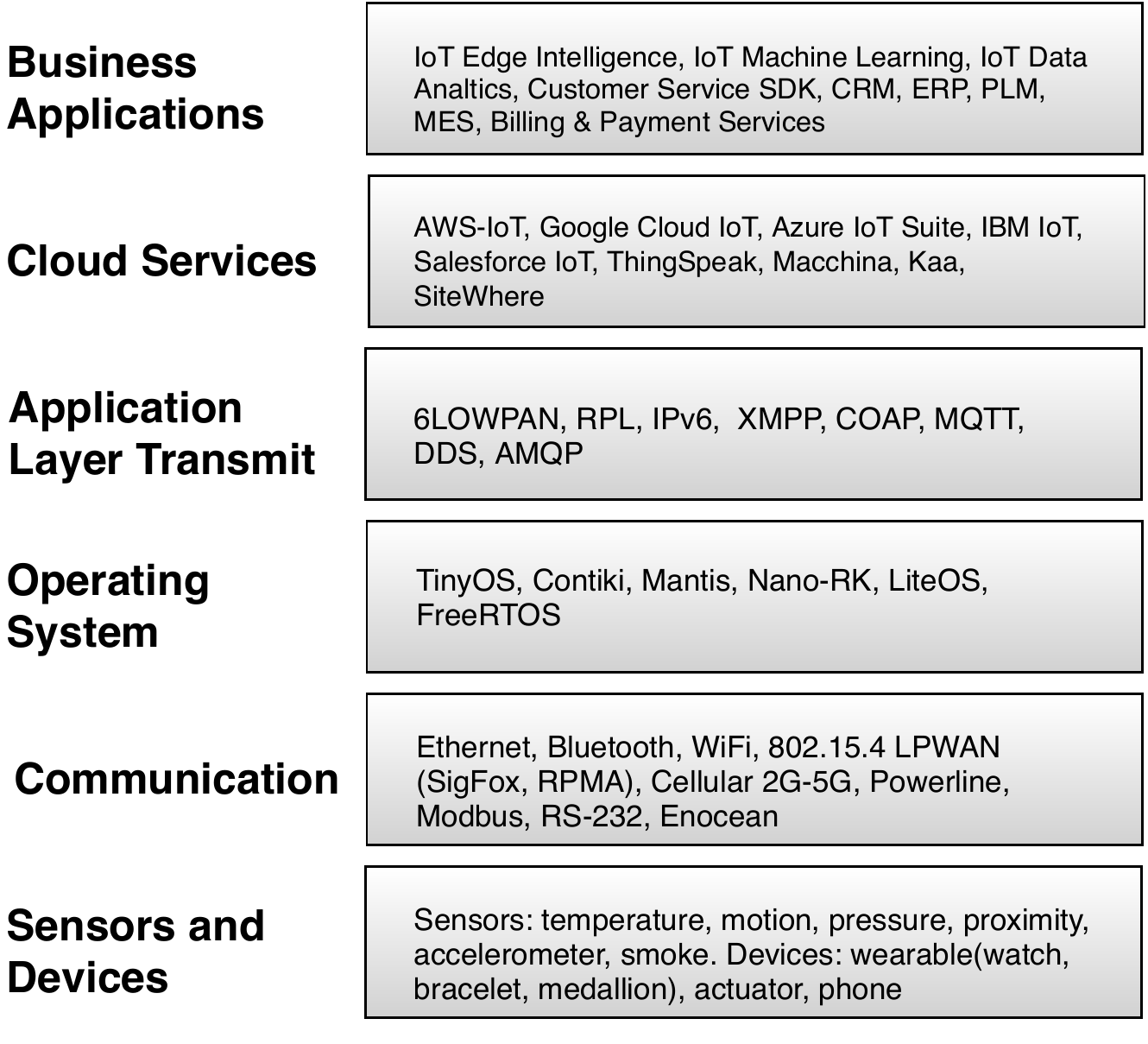}
    \caption{Technologies, devices and protocols used in an IoT System Implementation.}
    \label{fig:iottech}
\end{figure}

There are a variety of communication techniques, sensors and firmwares used by IoT devices. In the sensing layer, sensors are used to collect real time data from the surrounding environment such as temperature, humidity, motion etc. \cite{khan2012future} The end-device can actually be dumb or smart based on its data processing capability. While a smart phone has high computation power, an RFID tag provides an identification without any processing. The communication layer provides connectivity to other IoT devices or through a gateway. An IoT device may directly send data to the cloud using a cellular connection (4G, 5G) or through a gateway with short range communication such as Wifi, Bluetooth, or Zigbee. The IoT devices are deployed with embedded component-based operating systems or platforms that demand lower resources than common operating systems. The message is carried in various formats in the application layer such as Message Queue Telemetry Protocol (MQTT) or Constrained Application Protocol (COAP) to the cloud for analysis or storage. The data processing layer is generally located in the cloud through specific IoT services (e.g., AWS IoT, Google IoT, etc. ) whose job it is to analyze the collected IoT data and to make decisions based on the outcome. In some cases, historical data is also used in the data processing layer to improve decision making by utilizing machine learning techniques. This layer executes and presents the results of the data processing layer to applications to produce valuable services. The applications are commonly part of the enterprise level, which performs various tasks for users, customer relations, healthcare business, etc.

\subsection{IoT in Hospitality}

\begin{figure}[ht]
    \centering
    \includegraphics[scale=0.6]{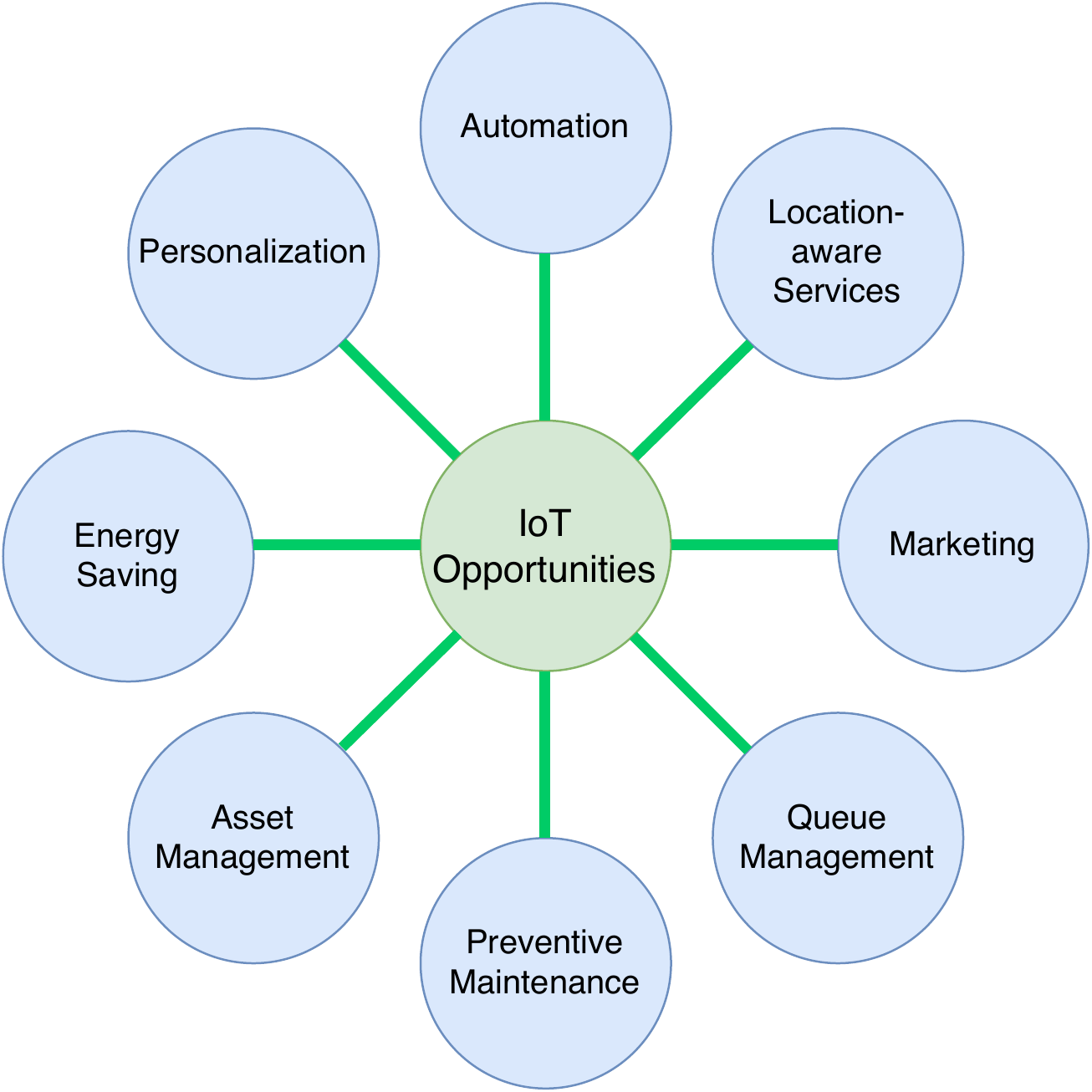}
    \caption{Hospitality sectors benefit IoT in many ways.}
    \label{fig:advantages}
\end{figure}

Hospitality is one of the pioneer domains that benefits from IoT in many ways \cite{car2019internet,pelet2019internet}. Improving customer satisfaction with novel services such as personalization increases the market share of the hospitality organization \cite{gretzel2015smart}. Intellectualization in service provision relies on historical data, user preferences and  accurate and timely data acquisition \cite{gretzel2018creating}. This has become possible with pervasive connected devices such as sensors, wearables and actuators \cite{xiang2017big}. Such devices enhance the organization’s interaction with consumers, and helps organizations better understand and respond to customer needs. IoT devices can also help improve efficiency for back of house management. Preventive maintenance, asset management and energy saving are some of the advantages afforded by IoT \cite{seventrend}, Fig. \ref{fig:advantages}. For instance, critical infrastructure, HVAC, and a freezer in the kitchen, may all be monitored using appropriate sensors, which will detect problems before they cause catastrophic failures. IoT devices may also inform staff of long lines in buffet restaurants or at amusement parks. Employing IoT devices can help increase safety, reduce costs, and decrease waiting time in a queue

\begin{figure}[ht]
    \centering
    \includegraphics[scale=0.7]{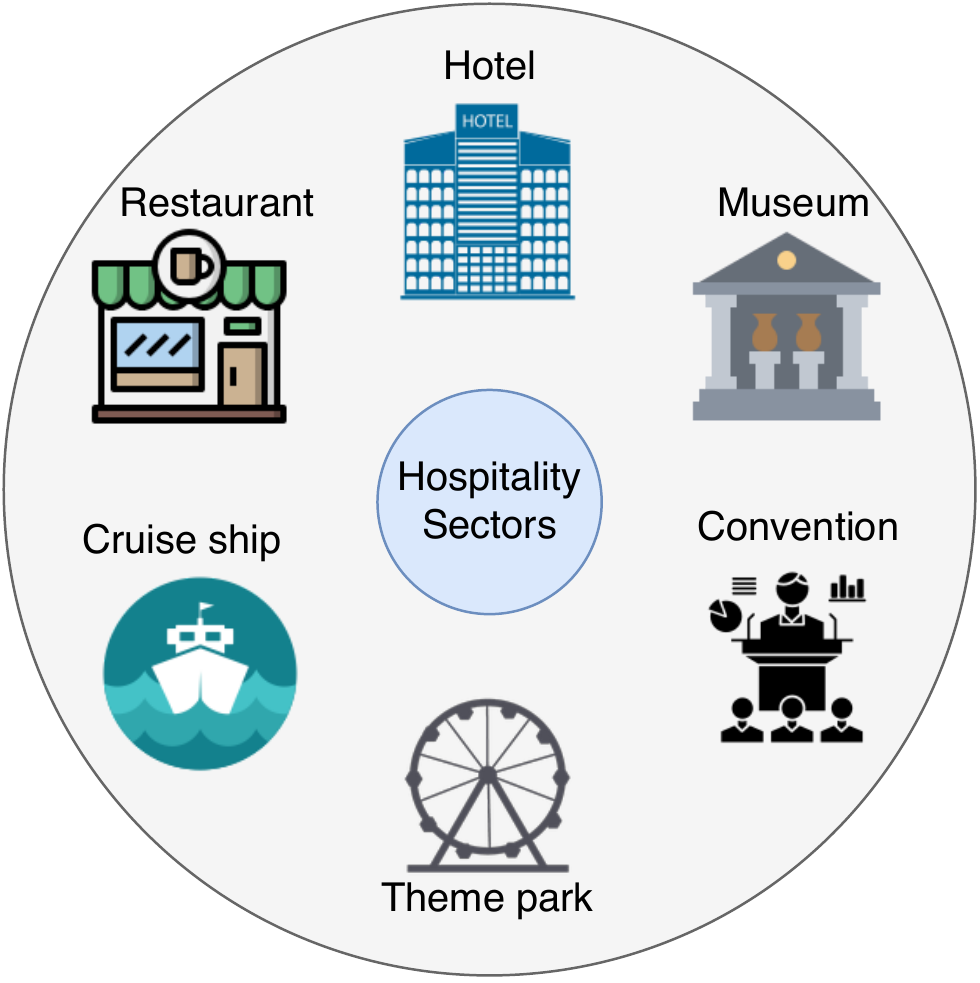}
    \caption{IoT is utilized in various hospitality sectors.}
    \label{fig:sectors}
\end{figure}

Each sector in the hospitality domain benefits from this trending technology as shown in Fig. \ref{fig:sectors}. 


Hotels are focusing on smart rooms with voice controlled items like curtains, TVs, and lights, that provide the ultimate comfort to guests \cite{marriott}. Self check-in via smart locks is another advantage offered in hotels and vacation rentals. Theme parks are utilizing this idea to smoothly handle large crowds of people efficiently in specified areas and at specific times. Disney uses the 'magicband' to streamline the visit, starting from the airport arrival all the way through payments in the park and checkout. Museums are creating smart interactivity with visitors according to their interests. They are also using IoT to protect sensitive collections using sensors by keeping the artwork in ideal temperature and element exposure conditions. Cruise ships are also utilizing IoT to manage operations on board. They can locate passengers and crew aboard the ship at all times. The Ocean Medallion, used by Carnival, accelerates the embarkation process and eases navigation on the ship. They enhance revenues by placement of products and merchandise for viewing based upon the history of guests' movements aboard the ship, provide real-time information to guests about events, exhibits, and dining, and avoid congestion through real-time location information. Restaurants, smart tourism, event management are some other sectors that employ IoT.

\subsection{Background on Security, Privacy and Ethics}
\begin{figure*}[ht]
    \centering
    \includegraphics[scale=0.8]{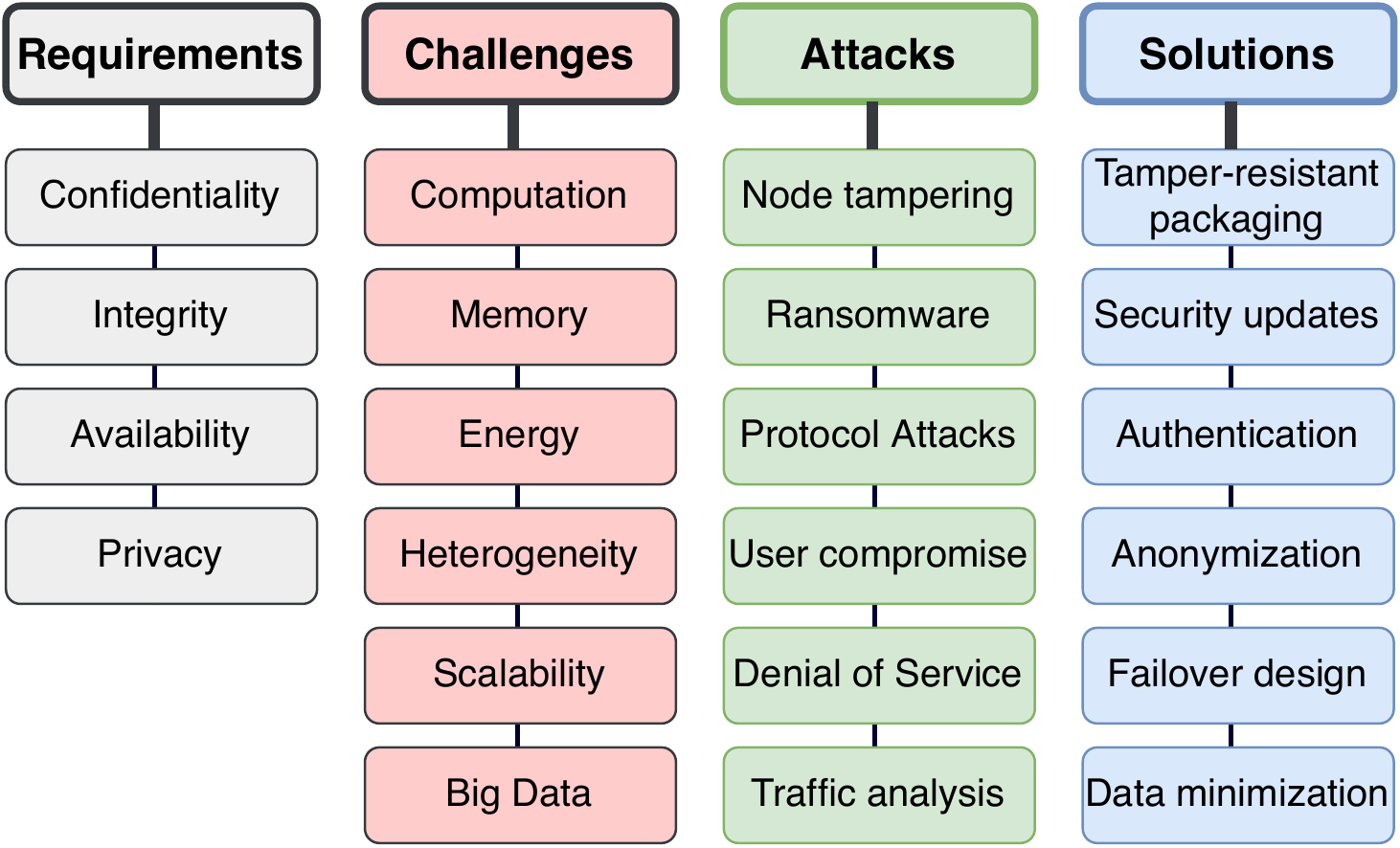}
    \caption{Requirements, Challenges, Attacks and Solutions.}
    \label{fig:challenges}
\end{figure*}

Three factors that must be considered when using IoT are security, privacy and ethics, which are interrelated. Fig. \ref{fig:challenges} lists some high level security requirements, along with challenges and attacks, and potential solutions.

\textit{Security}: A system is considered secure if it holds three primary objectives; confidentiality, integrity and availability. Integrity is keeping the message intact between the sender and receiver. Confidentiality means that the message is not known by unauthorized users, and availability is the continuation of a service against disruption attacks.

\textit{Privacy}: Privacy is the right of a person or an entity to decide what type of information should be known by others, and to what extent. An individual should have control over the collection, processing, dissemination and subsequent use of data collected about him. While the security of the system and data is a business owner's responsibility, attacks to system security may cause problems like privacy leakage, which is out of the operator's control. Moreover, the system owner or third party services used for storage or processing may be a threat to a user privacy if they don't behave ethically.

\textit{Ethics}: Ethics studies "what is good and bad" or "what is right and wrong", and goes beyond this by including compliance with regulations. Any new technology designed for the benefit of people may also have negative side effects on individuals and society. Thus, it is imperative to define ethical rules and legal regulations to protect the individuals and society. In an IoT environment, since personal data will be in system owner's hand, and it may not be possible to control each data flow, thus ethical manner and observing user rights is highly significant. Regulations are evolving as regulators’ understanding increases as to IoT vulnerabilities. There are few statutes and regulations specifically for IoT applications presently, so laws of privacy, product liability, warranties and negligence are applied without a comprehensive view \cite{williams2014privacy,crootof2019internet,o2015internet}. In the absence of governmental regulation, the private managers of the technology must engage in responsible self-regulation. The European Union, through the General Data Privacy Act (GDPA), and the United States Federal Trade Commission (FTC) have attempted to apply existing regulations to IoT users’ data privacy, but without a coherent framework \cite{posadas2018internet,murphy2017recent}.

\section{Security for IoT in Hospitality}

Security is one of the most important factors hindering the wide adoption of IoT as the attacks have significantly increased. The attack surface is increasing with exponential rise in the number, complexity, and heterogeneity of IoT devices \cite{hossain2015towards}. There exists many active and passive attacks at various levels either targeting the device, communication infrastructure or the server. Social engineering, DoS attacks, man-in-the-middle attacks, user account, hardware and software compromise, traffic analysis, side channel attacks, malicious insiders are just few examples to name \cite{butun2019security}. Below are listed some security risks that might directly affect the operation in a hospitality environment.  

\textbf{Exploitation of Integrated Systems}: Any of the CIA (Confidentiality, Integrity and Availability) triad can be the target of a malicious activity which can come from inside or outside of a company. IoT devices in hospitality domain are deployed as part of a complete system interconnected via various communication techniques such as Bluetooth, RFID, WiFi. Therefore, one compromised device may make the whole system vulnerable to severe damages. For instance, a major casino in Las Vegas was hacked through a smart thermometer in the lobby fish tank \cite{fishtank}. The hacker infiltrated the network and accessed the database without being caught by traditional intrusion detection systems. This demonstrates the level of risk posed by IoT devices in the hospitality domain. 

\textbf{Ransomware}: Ransomwares could be a nightmare for a businesses in hospitality that disrupts the routine operation of the company that compels the owners to pay the amount requested by the hackers as the loss could be higher in terms of money and reputation. This has been a major issue for hotels in recent years. For example, in the case of Romantik See hotel in Austria whose locking system was frozen by a ransomware, the attack affected the computer system and prevented the employees from programming the keycards which caused customers not to be able to enter their rooms \cite{romantikhotel}. The hotel had to pay \$1800 worth of Bitcoin to end the attack.

\textbf{Denial of Service}: IoT devices have been the target of hackers to be used as bots to launch DDoS attacks as in the case of Mirai botnet \cite{antonakakis2017understanding}, which exploited IoT cameras. The hackers exploited the fact that most people do not change the default username and password. Thus, it is critical to set a qualified password as it is one of the common weak points exploited by attackers. Hospitality industry becomes a favorite target for this type of attacks since there are a lot of devices hosted. This sector has seen a great increase in attacks after finance \cite{doshospitality}. 

\textbf{Heterogeneity and Interoperability}: Variety in device type and lack of standards is another challenge that makes it difficult to develop a comprehensive security solution. Each device might use different set of communication standards and security settings that makes harder to control the overall system. For instance, in a hotel room, a TV communicates through WiFi, coffeemaker is using bluetooth, a smartlock sends messages over Zigbee, and RFID is used inventory management. Thus, it is hard to come up with a unique interoperable solution that works for all devices and protocols.

\textbf{Physical Attacks}: Physical attacks are easier to achieve in a hospitality environments as guests have physical access to devices such as in a hotel room. Node tampering, malicious node injection, code injection are some attack types that can be pursued easier with physical access to devices \cite{hossain2015towards}. The attacker may capture the device, extract cryptographic keys, modify the programs, replace the device with a malicious one \cite{atlam2020iot}. Thus, devices should have tamper-resistant packaging. There have been incidents 
where perpetrators posed themselves as guests by plugging their own device into unsecured IoT devices in the lobbies to hack into the system. Even the devices may be counterfeit that may contain malicious codes. To mitigates such attacks, some chips contain secure certificates (physically unclonable function) that are physically and electronically protected \cite{maple2017security}. These chips can be used to countermeasure physical attacks, but it will increase the cost significantly. In order to minimize the risk from customers, guest network traffic should be separated from rest using virtual networks or SDN in the network architecture. This will prevent hackers to exploit an infected device that a customer may inadvertently connect to network.

\textbf{Security and Software Updates}: Deployed IoT devices are intended to run for long years and attacks might change in the meantime. Security and other software updates are important measures for system protection. However, updating and patching the operating system on these devices is not straightforward. They also do not let to run third party security solutions. The main challenge is to have an automated mechanism to find out any updates, and then downloading them securely from their authentic resources. In many cases, hospitality personnel are not technical and they ignore these updates. As a result some systems might be using former insecure versions of some standards (i.e., TLS1.1 which is not secure instead of TLS1.3). 




\section{Privacy for IoT in Hospitality}

Pervasive data collection with omnipresence of IoT devices and ultraconnectivity provides quantitative and qualitative private information about an individual in many hospitality applications. Moreover, data collection is passive and non-intrusive because of seamless interaction, thus, people are mostly unaware of being tracked. This weakens privacy perception and allows extensive data collection. The collected data may reveal information about a person and his activities with time, location and context details which can be used to do targeted advertisement on individuals. Thus, IoT poses a bigger threat to privacy of individual when compared to the Internet. Attacks to privacy might come from various sources such as an insider employee, from another customer as well as a result of a security breach and misuse of personal information by the businesses. Some of the well known privacy attacks are identification, tracking, profiling and linkage \cite{ziegeldorf2014privacy}. As IoT usage grows rapidly in hospitality, privacy is the most contentious issue from the consumer perspective. Below, we summarize specific privacy threats within the Hospitality domain: 

\textbf{Tracking}: In hospitality, personalized, location-based and context-aware service is key to improved customer satisfaction which can be pursued by historical data analysis and continuous connectivity. This requires collection and preservation of consumer data by tracking a person's physical location through available IoT devices such as wearables (i.e., Carnival's Medallion or Disney' wristbrand \cite{carnival}). In order to make accurate recommendation to users, which will increase the revenue, business owners aggressively collect such data. For instance, on a cruise ship or in a theme park, there are attractions going on that users may want to get notified. A guest may be offered his favorite drink while passing through a bar. Some visitors may be offered promotional foods to manage the queue for popular activities. However, tracking, monitoring and reporting of the users’ actions will reveal behavioural patterns and leak sensitive information \cite{kroger2018unexpected}. For example, eating habits or fitness tracking can tell health status which may be used by an health insurance company to determine the quote or even deny in unauthorized way. 
This type of information is also important to profile the individuals that can be exploited for marketing by third parties if they can access.

\textbf{Profiling}: Aggregation and correlation of information dossiers can create detailed profile of people (virtual identity) and help de-anonymize sensitive data. A person may be classified according to some criteria using data trails left behind. Profiling can lead to price discrimination, unsolicited advertisement and social engineering. For instance, likes and dislikes can be inferred from the activities during the day which then can be shared with other companies for marketing purposes. 
Even if the data is kept anonymous, it can still be used for market and price discrimination. Moreover, linking pieces from different datasets is one of the major privacy concerns \cite{ziegeldorf2014privacy}. A piece of data which is not identifiable in isolation may actually be personal data when combined with others, 
\textit{quasi-identifier}. This type of threat is valid especially when multiple data from different sources such as Facebook are merged.

\textbf{Identification}: Identification is relating an identifier (name, address) with a private data. Surveillance cameras are intensively used in hospitality environments for monitoring and managing the crowd in addition to security purposes. Moreover, facial recognition capability of cameras are exploited for marketing. Data from cameras and voice-enabled devices especially when combined with other databases could be big threat to personal privacy as it reveals clear information about individuals and can be linked to the real identity of a person.  In order to prevent such attacks, instead of using real identity of a person when storing information, \textit{pseudonyms} could be used. A pseudonym hides the real identity by creating a random identifier. For instance, as vacation rental owner, you may want to know if the guest is present, or number of guests in the property so that you can send the housekeeping staff or to let the next guest to check in. Since using a camera will be a privacy violation, alternative methods that keeps the guest identities unrevealed such as noise-monitoring, CO2 sensors and occupancy detectors are helpful to manage the property.

\textbf{Compromised Device}: There is a lot of IoT devices used in the Hospitality domain. For instance, voice-controlled devices are key to create smart rooms in hotels. The user can control the items in the room with voice commands which provides great comfort to the guest. These devices are always listening to capture commands regardless of the input whether it is just a regular conversation or a specific command to the device. The user never knows what is happening in the device. It might be performing unwanted functions or collecting data more than it needs. Furthermore, the device may not have been designed to do so, but it may be compromised using some techniques mentioned in the previous section, then it raises the most critical risk especially if it is equipped with a camera. Also, since same devices are used by many people in hospitality environment, malicious customers could be threat to others. For example, a customer loads his profile to the TV in the hotel room, and the next person using the same room might learn previous customers' profile information or watch whatever was in the history if it is not deleted. 

\textbf{Traffic Analysis}: Even though the device is running properly, the data may be intercepted during transmission. This type of attack is countered with end-to-end data encryption. But, this does not solve the problem completely. Even if the data is encrypted, traffic analysis on encrypted data, cryptanalysis attacks or side channels might still be used to gather some information \cite{coull2014traffic}. For instance, by looking at running time and duration of smart devices in hotel room (coffee maker, shower), some information can be inferred about the guests such as number of people in the room. This type of attacks exist in other domains using similar devices like smart meter (i.e., power consumption reveals types of devices used). 

\textbf{Data Processing on cloud}: In a typical IoT system, data is sent to cloud server for data analysis which puts the data on a third party platform. This introduces another type of risk since the control over the data is handed over to service provider \cite{henze2016comprehensive}. The main concern with this type of data processing is that the users may not be sure how the data will be used because of lack of transparency. Homomorphic encyription, computation on encrypted data, is utilized to address this problem, but it only provides limited computation \cite{gentry2009fully}. Client side personalization, data perturbation, obfuscation, anonymization are some current techniques to improve privacy \cite{spiekermann2008engineering}.

\section{Ethics for IoT in Hospitality}

Convenience and privacy/security in the IoT context, seemingly two contradicting feature, is difficult to achieve simultaneously, but they could be reconciled to some extent by better understanding customer needs, applying best security and privacy practices, and abiding by rules and regulations.  
However, this may not be enough as security and privacy regulations may not always be quickly enforced by law. This brings the discussion to ethics which means Hospitality stakeholders sometimes would need to make decisions on their own considering what is right or wrong as the law/regulation/compliance may not be applicable or may not even exist. Indeed, ethics often lags behind technological developments, thus it is essential to identify ethical issues to consider while implementing and running IoT systems to avoid disastrous consequences. In this regard, using IoT in hotel or similar environment requires intensive data collection, thus it has ethical responsibilities for consumer, business owner and employees. These issues are often neglected as the cases were not common in the past. However, with the increasing deployment of IoT devices in Hospitality domain, there needs to be a discussion and awareness on this issue.  Below, we list issues regarding ethical considerations of IoT use in the hospitality domain:

\textbf{User Privacy Agreement}: 
Privacy notion is subjective and greatly depends on individual perception \cite{ziegeldorf2014privacy}. Each person might request a different level of privacy, even the same person might desire various levels depending on time, context and location. For instance, privacy expectation at home is not same as in public. Thus companies manufacturing IoT devices and hotels using these devices should recognize range of privacy expectations of users. The services and information collection should be customizable in addition to opt-in and opt-out flexibility. However, in a typical "end user licence agreement" (EULA), consumers are not offered a second alternative than approving the whole document, but they are enforced to click to advance and get the service. Guest forms requiring “notice and consent” as to the terms and conditions of an agreement are arguably entered into under “duress” since they are generally nonnegotiable and not understandable to the guest. Moreover, this document is either too detailed for the customer to go through and understand, or it is just a summary and not very informative.  Although terms such as choice of venue and law have been validated by courts, waivers as to privacy and personal data take this to another level not yet decided by legal precedent \cite{blanke2017privacy}. Alternatives to “notice and consent” model should be developed.

\textbf{Data collection and usage}: Data management in this context is the greatest problem to be addressed very well. The data will probably provide demographic information that may be used for profiling and marketing purposes which may not be user's choice. Who has the property right on the collected data? If the users own the data others created about them, then they should be able to control it. However, the method for data acquisition, length of retention and sharing policy is not always clear to customers. The user mostly has incomplete knowledge about the consequences of data release \cite{baldini2018ethical}. As guest data is transmitted to Artificial Intelligence (AI) for further processing, there is potential for discriminatory results which would violate American civil rights laws \cite{coglianese2016regulating}. Even though the consumer is asked for consent about using a device, it is not realistic to ask for every possible type of IoT device during the day \cite{allhoff2018internet}. Most of the time, data is gathered without an active participation and even without noticing. Lack of  communication interface in IoT devices limits the capability of changing security settings similar to the case where you can turn on and off history on a browser. Once the data is collected, it is the business owner's responsibility to protect it. Even though the users are asked for consent, they can not fully grasp all details and future cases to make a decision. Moreover, the immediate access to a benefit might cause ignorance of long term negative impact. Thus, the unintended data should not be used beyond its purpose. 

Another aspect of data usage is, it could be accessed by legal authorities. One U.S. Supreme Court case which reaches a tangential issue decided that it is not an illegal search for police to obtain data on cellphone tower usage by a suspect. Does this open the door to use of IoT data for private purposes as not within personal privacy protection \cite{ohm2018many}? In this context, another aspect might be about changing laws when a tourist is traveling to other countries. For instance, European Union (EU) enacted a new law called The General Data Protection Regulation (GDPR) on data protection and privacy, which also regulates the use of their citizen's data in other countries. However, a tourist from a non-EU country traveling to Disney in the US, where his/her data is collected with Disney's wristband will not be subject to GDPR. It will then be totally up to Disney how to manage this data other than just applying US privacy protection laws.


\textbf{Responsibility Ascription}: IoT has impact in physical realm as the decisions may depend on the data from these devices. These devices may interfere with human activities spontaneously \cite{tzafestas2018ethics}. Although IoT in hospitality is mostly used for service recommendation to customers, it is also used to automate critical infrastructures and some processes such as kitchen management. Automation may cause unexpected behaviours and malfunctioning because of hardware problems or algorithmic mistakes. This raises the question who will be legally responsible for potential damages. For instance, wrong sensor readings in freezer or cookers may endanger food safety. Manufacturing, programming and controlling of these devices overseas further complicates the determination of responsibility.

\textbf{Changing Skill Set}: Emerging technology requires a shift in the skill set to land a proper job, thus current employees may get frustrated about unemployment or not being able to adapt to new world. They may need to change the way they are working which can create anxiety and lead to resistance. Trending technologies such as IoT influence companies to decide new hires and layoffs. The companies need to address this type of concerns by investing in education of current employees. Obviously, this is an ethics question which needs to depend on the tradeoff between costs and investing on loyal employees. 


\textbf{Risk Management Cost}: Tackling with security and privacy may incur extra cost in addition to technology investment cost. For instance, applying security updates is not easy for IoT devices which may require replacement with new ones. The management should make a decision on security level and additional cost induced \cite{abobakr2017iot} and find an optimum point. Additionally, dishonest operators that collects, use, even share data with others may have some advantage over honest business owners in the short term. 




\section{General Guidelines}
This section presents general guidelines that can help avoid potential problems in using IoT in hospitality environments. 

\textit{Hardware Security}: Using original devices manufactured by known companies will reduce risk of being attacked through compromised devices. Using hardware that incorporates security measures will improve the integrity such as tamper detection, secure execution environment.

\textit{Security by design}: Security and privacy should be incorporated into design and development from the beginning including device manufacturing, system design, deployment and running. This approach is important because it is difficult to detect and fix errors after implementation before any incident. IoT implementation should adopt a holistic approach which considers the whole lifecycle by integrating security in all levels ranging from design to deployment.

\textit{Failover design}: IoT devices especially critical part of a system should continue working when Internet connection is off. For instance door lock should still be opened with an alternative method.

\textit{Firmware Updates}: 
Device firmwares must be updated regularly to apply available patches. Automatic update mechanism could be useful while checking for the existence of updates at frequent intervals.

\textit{Data management}: 
Collect data that has only immediate use not for future use and delete it as soon as possible
minimize the data collected and retained, and dispose it when you do not need it. Larger data will attract more attackers. Collected data should not expose individuals to any type of discrimination.

\textit{Policy}: 
Security policy should be defined such as the use of personal devices for work purposes where compromised devices could be exploited. Attack prevention, detection and mitigation techniques should be determined.

\textit{Education}: 
Education on contemporary technology is important for employees to gain new skill to adapt to the new technology to be marketable. Cyberliteracy such as setting a strong password (simple but effective) will help secure the system.

\textit{Informed consent}: 
Consumers must be clearly communicated in an understandable language and maximum transparency should be provided. The user should indicate his/her wish by a clear affirmative action \cite{allhoff2018internet}. They should also be able to manage their data.



\section{Conclusion}

Hospitality is one of pioneer domains utilizing IoT to improve customer satisfaction and operational efficiency whereas the threat to privacy and security will continue to grow with the further increase in the number, variety, accuracy and adoption of these devices while hospitality environments pose specific challenges. In this paper, we explored security, privacy and ethical concerns regarding the use of IoT applications in hospitality industry.
Consumers need to be aware of privacy concerns and business owners should act responsibly by implementing sound security measures and abiding by ethical rules and legal regulations. Data which is legally available to the systems’ owner is ethically restricted to its intended use and not for improper purposes.  Even if not legally mandated, and even if a waiver of privacy and confidentiality has been signed by the user, the owner should be bound ethically not to use data for improper purposes. 

%
%
%
\bibliographystyle{IEEEtran}
%

\bibliography{References.bib}






\end{document}